\documentclass[aps,preprint,floatfix,nofootinbib,showpacs]{revtex4-1}
\pdfoutput=1
\usepackage{graphicx,color,float, amsmath}
\usepackage{hyperref}
\usepackage{MnSymbol}%
\usepackage{wasysym}%
\usepackage{color}
\newcommand{\real}{\Re {\rm e}}

\begin{document}


\title{Higgs boson pair productions in the Georgi-Machacek model at the LHC}

\def\slash#1{#1\!\!/}

\renewcommand{\thefootnote}{\arabic{footnote}}

\author{
Jung Chang$^1$, Chuan-Ren Chen$^2$, Cheng-Wei Chiang$^{3,4,1}$}
\affiliation{
$^1$Physics Division, National Center for Theoretical Sciences, Hsinchu, Taiwan 30013, R.O.C.\\
$^2$Department of Physics, National Taiwan Normal University, Taipei, Taiwan 11677, R.O.C.\\
$^3$Department of Physics, National Taiwan University, Taipei, Taiwan 10617, R.O.C.\\
$^4$Institute of Physics, Academia Sinica, Taipei, Taiwan 11529, R.O.C.
}
\date{\today}

\begin{abstract}
Higgs bosons pair production is well known for its sensitivity to probing the sign and size of Higgs boson self coupling, providing a way to determine whether there is an extended Higgs sector.  The Georgi-Machacek (GM) model extends the Standard Model (SM) with an $SU(2)_L$ triplet scalar field that has one real and one complex components.  
The Higgs self coupling now has a wider range than that in the SM, with even the possibility of a sign flip.
The new heavy singlet Higgs boson $H^{0}_{1}$ can contribute to s-channel production of the $hh$ pairs.  In this work, we study non-resonant/resonant Higgs boson pair productions $p p \rightarrow hh$ and $p p \rightarrow H^{0}_{1} \rightarrow hh$, focusing exclusively on the contribution of $H^0_1$. 
We show the sensitivity for Higgs boson pair production searches at the 13-TeV LHC with the luminosities of $3.2,\ 30$ and $100$~fb$^{-1}$.
\end{abstract}

\maketitle

\section{Introduction}

After the discovery of Higgs boson at the LHC \cite{atlas, cms}, couplings of the Higgs boson to certain other Standard Model (SM) particles have been measured and the best fit is performed with the result very close to the SM expectation~\cite{higgs}.  However, the Higgs boson self coupling, a key parameter to test the structure of Higgs potential and electroweak symmetry breaking, has not yet been measured.  At the LHC, Higgs boson pair production is known to be the primary process where one can use to determine this coupling~\cite{Djouadi:1999rca,baglio, ggf_hh, hh-sm, bbaa_hh, 4b_hh}.  Nonetheless, it is expected to be a challenging measurement due to its low production cross section predicted in the SM, $\sigma(p p \rightarrow h h)_{\rm SM} \sim 40~ \rm{fb}$ at the 14-TeV LHC~\cite{LHCXS, deFlorian:2016spz, Borowka:2016ehy, plehn}.  In the SM, tree-level Higgs trilinear and quartic self couplings are given as 
\begin{eqnarray}
 g_{hhh}^{\rm SM} = \frac{3m_h^2}{v} ~,~~  g_{hhhh}^{\rm SM} = \frac{3m_h^2}{ v^2} ~,
 \label{g_SM}
\end{eqnarray}
where $m_h$ is the Higgs boson mass, and are related by a factor of the vacuum expectation value (VEV) $v = 246~ \rm{ GeV}$.

Physics beyond the SM (BSM) can easily affect the Higgs pair production cross section at the LHC through either modification in the top Yukawa coupling and/or new colored particles running in the triangle and box loops (non-resonance effects), or the existence of new heavy scalars decaying into Higgs pairs (resonance effect).  The enhancement in production cross section can reach a few orders of magnitude in some cases~\cite{hh-bey, hh-susy, Godunov:2014waa,chen_low, hh-mi, hh-jc}.
Currently, the ATLAS and CMS Collaborations have imposed upper limits on the production cross section ($bb\gamma\gamma$) and production cross section times branching ratios ($4b$, $\gamma\gamma W W^*$ and $\tau\tau bb$) with various categories of signal final states in Higgs pair searches at the 13-TeV LHC~\cite{atlas_8, cms_8, atlas_13, atlas_13_2, atlas_13_3, cms_13, cms_13_2}: $3.9 ~\rm{pb}$, $330~\rm{ fb}$,   $25~\rm{ pb}$ and $508~\rm{ fb}$ for the $\gamma\gamma bb$,  $4b$, $\gamma\gamma W W^*$ and $\tau\tau bb$ channels, respectively.

The Georgi-Machacek (GM) model, proposed in the mid 1980s~\cite{gm1, gm2}, provides a good way to generate Majorana mass for neutrinos through the type-II seesaw mechanism while preserving the custodial symmetry at tree level.  In addition to the SM-like Higgs boson $h$, the extended Higgs sector has another three neutral scalars, among which two are CP-even ($H^0_1$ and $H^0_5$) while the other is CP-odd ($H^0_3$), where the subscripts denotes their representations under $SU(2)_L$.  One distinctive feature of this model is that the couplings between $h$ and the SM weak gauge bosons, $g_{hVV}$, can be larger than their SM values.  Phenomenology of this and similar models, including their supersymmetric and dark matter extensions, at both hadron and lepton colliders have been extensively studied~\cite{Gunion:1989ci,Gunion:1990dt,Haber:1999zh,Aoki:2007ah,Godfrey:2010qb,Logan:2010en,Falkowski:2012vh,Chiang:2012cn,Englert:2013zpa,Englert:2013wga,Chiang:2013rua,Hartling:2014zca,Chiang:2014hia,Chiang:2014bia,Godunov:2014waa,Hartling:2014aga,Chiang:2015kka,Godunov:2015lea,Chang:2003zn,Cort:2013foa,Garcia-Pepin:2014yfa,Chiang:2015rva,Chiang:2015amq,Campbell:2016zbp}.

With the GM scalars also in the Higgs potential, the SM-like Higgs trilinear coupling and its couplings to the SM fermions are modified, with the possibility of enhancing the non-resonant Higgs boson pair production cross section.  Furthermore, $H^0_1$ can also mediate the Higgs boson pair production, and virtually the $gg\to H^{0}_{1}\to hh$ channel dominates at the LHC when $H^{0}_{1}$ can be produced on shell.

Constraints on the GM model have already been studied from unitarity of scalar field scattering amplitudes, tree-level stability of the Higgs potential, and Higgs boson precision measurements~\cite{Hartling:2014zca,Chiang:2012cn, Chiang:2015amq, Chiang:2015kka}.
The most stringent constraint allows only a small window in the interaction between the Higgs boson and weak gauge bosons $\kappa_V \equiv g_{hWW}/g_{hWW}^{\rm SM} = 0.94^{+0.11}_{-0.12}$~\cite{higgcision}. 
Ref.~\cite{Chiang:2015amq} studied the constraints on the $\alpha$-$v_\Delta$ plane using a $\chi^2$ fit to the data of Higgs boson production at LHC Run-I, including both gluon-gluon fusion (GGF) and vector boson fusion processes with the tree-dominated $b\bar{b},\ \tau^+\tau^-, ZZ$ and $WW$ decay channels.  Within the $2\sigma$ contour, the mixing angle $\alpha$ and the VEV of the Higgs triplet field $v_\Delta$ are found to roughly fall within the following ranges: $-50^\circ\lesssim \alpha \lesssim40^\circ$ and $0 \le v_\Delta \alt 50$~GeV, as shown explicitly in Fig.~1 of Ref.~\cite{Chiang:2015amq}. 
In this work, we will focus on the 125-GeV Higgs boson pair production via the non-resonant $p p \rightarrow h h$ channel and the resonant $p p \rightarrow H^{0}_{1} \rightarrow h h$ channel in GM model.

The rest of this paper is organized as follows.  In the Section~\ref{sec:model}, we review the GM model and show the relevant couplings.  The pair production of Higgs bosons in the model is discussed in Section~\ref{sec:hhprod}.  Section~\ref{sec:result} shows our numerical results and direct search constraints from the 13-TeV LHC.  Finally, we give a summary of our work in Section~\ref{sec:con}. 
%

\section{Georgi-Machacek model}
 \label{sec:model}

In the GM model, two $SU(2)_L$ triplet scalar fields, $\chi$ with hypercharge $Y=1$ and $\xi$ with $Y=0$, are introduced to the Higgs sector in addition to the $SU(2)_L$ doublet $\Phi$ with $Y=1/2$ already in the SM.  In this paper, we use the convention that $Q=T_3+Y$ with $Q$ and $T_3$ being the electric charge and the third component of the weak isospin, respectively.
Writing in an $SU(2)_L\times SU(2)_R$ covariant form, we have
\begin{eqnarray}
\Phi = \begin{pmatrix}\phi^{0*} & \phi^+ \\ -(\phi^+)^* & \phi^0\end{pmatrix},\ \Delta = \begin{pmatrix} \chi^{0*}&\xi^+&\chi^{++} \\ -(\chi^+)^* & \xi^0 & \chi^+ \\ (\chi^{++})^* & -(\xi^+)^* & \chi^0\end{pmatrix},
\end{eqnarray}
where we use the following phase convention for the scalar field components: $\phi^- = (\phi^+)^*,\chi^{--} = (\chi^{++})^*, \chi^- = (\chi^+)^*, \xi^- = (\xi^+)^*$.
As in the SM, due to the instability of the Higgs potential, the neutral component of $\Phi$ spontaneously develops a VEV to break the electroweak symmetry and to induce VEVs for the neutral components of $\Delta$.
We can parameterise these neutral fields as
\begin{eqnarray}
\phi^0={1\over\sqrt{2} } (v_\phi+\phi_r+i\phi_i) ~,~ \chi^0 = v_\chi + {1\over\sqrt{2} }(\chi_r + i\chi_i) ~,~ \xi^0 = v_\xi + \xi_r ~,
\end{eqnarray}
where $v_\phi$, $v_\chi$ and $v_\xi$ denote the VEVs of $\phi$, $\chi$ and $\xi$, respectively.  In the case of vacuum alignment $v_\chi = v_\xi \equiv v_\Delta$, we have $v^2\equiv v^2_\phi + 8 v^2_\Delta = (246~\mbox{GeV})^2$, and define $\tan \beta \equiv v_\phi / (2\sqrt{2} v_\Delta)$.
More explicitly, the Higgs potential in the GM model is given by
\begin{eqnarray}
V(\Phi,\Delta) = 
&& 
{1\over 2} m^2_1 {\rm tr} [\Phi^\dagger \Phi]+{1\over 2} m^2_2 {\rm tr} [\Delta^\dagger \Delta]
+ \lambda_1 ({\rm tr} [\Phi^\dagger \Phi])^2+\lambda_2 (\rm{tr} [\Delta^\dagger \Delta])^2
\nonumber \\
&&
+ \lambda_3 {\rm tr} [(\Delta^\dagger \Delta)^2]+\lambda_4 \rm{tr} [\Phi^\dagger \Phi]\rm{tr} [\Delta^\dagger \Delta]
+ \lambda_5 {\rm tr} \left[ \Phi^\dagger {\sigma^a\over 2}\Phi {\sigma^b\over 2} \right]
\rm{tr} \left[ \Delta^\dagger T^a\Delta T^b \right]
\nonumber\\
&&
+ \mu_1 {\rm tr} \left[ \Phi^\dagger {\sigma^a\over 2}\Phi {\sigma^b\over 2} \right] 
(P^\dagger \Delta P)_{ab}
+ \mu_2 {\rm tr} \left[ \Delta^\dagger T^a\Delta T^b \right] (P^\dagger \Delta P)_{ab} ~,
\end{eqnarray}
where $\sigma$'s and $T$'s are the $2\times 2$ and $3\times 3$ matrix representations of the $SU(2)$ generators, and
\begin{align}
P = 
\frac{1}{\sqrt2}
\begin{pmatrix}
-1 & i & 0 
\\
0 & 0 & \sqrt2
\\
1 & i & 0
\end{pmatrix} ~.
\end{align}

After the $SU(2)_L\times SU(2)_R$ symmetry is broken down to the diagonal $SU(2)_L$, the scalar fields in the GM model can be classified into different representations under the custodial symmetry transformation: $\Phi$ is decomposed into a $\bf 3$-plet and a singlet and $\Delta$ into a $\bf 5$-plet, a $\bf 3$-plet and a singlet.  Among the neutral fields, we have two CP-even singlets $H_\Phi^1=\phi_r$ and $H_\Delta^1=\sqrt{1/ 3}\xi_r + \sqrt{2/3} \chi_r$ that mix through a mixing angle $\alpha$ to render two physical Higgs bosons:
\begin{eqnarray}
h = \cos\alpha H_\Phi^1-\sin\alpha H_\Delta^1,\ \ \ H_1^0 = \sin\alpha H_\Phi^1 + \cos\alpha H_\Delta^1
~,
\end{eqnarray}
and one CP-even $H^0_5$ given by
\begin{eqnarray}
H^0_5 = \sqrt{1\over 3} \chi_r - \sqrt{2\over 3} \xi_r ~.
\end{eqnarray}
Here, we take $h$ to be the SM-like Higgs boson of mass $125$~GeV.
The two CP-odd $\bf 3$-plet fields mix via a mixing angle $\beta$ to produce a physical $H^0_3 = -\cos\beta \phi_i + \sin\beta \chi_i$ and a Goldstone boson that becomes the longitudinal component of the $Z$ boson.  Because of the custodial symmetry, the different charged states within each representation are almost degenerate in mass, subject to small mass splitting $\sim {\cal O}(100)$~MeV due to electromagnetic corrections.  In the following, we will ignore such small mass differences and denote the Higgs masses by $m_{H_5}$, $m_{H_3}$, $m_{H_1}$, and $m_h$ for the physical $\bf 5$-plet, $\bf 3$-plet, heavy singlet, and SM-like Higgs boson.

The five dimensionless scalar couplings $\lambda_1-\lambda_5$ in the GM model can be expressed in terms of the physical Higgs masses and the mixing angles $\alpha$ and $\beta$ as
 \begin{eqnarray}
&& \lambda_1 = {1\over 8v^2s^2_\beta}
\left( m^2_hc^2_\alpha + m^2_{H_1^0}s^2_\alpha \right),
\nonumber \\
&& \lambda_2 = {1\over 6v^2c^2_\beta}
\left[ 2m^2_{H_1^0}c^2_\alpha+2m^2_h s^2_\alpha+3M_2^2-2m^2_{H^0_5}+6s^2_\beta(m^2_{H^0_3}-M^2_1) \right],
\nonumber\\
&& \lambda_3 = {1\over v^2 c^2_\beta} \left[ s^2_\beta \left( 2M^2_1-3m^2_{H^0_3} \right) 
+ m^2_{H^0_5}-M^2_2 \right],
\nonumber \\
&& \lambda_4 = {1\over 6v^2c_\beta s_\beta} \left[ \sqrt{6} s_{\alpha} c_\alpha \left( m^2_h-m^2_{H_1^0} \right) + 3c_\beta s_\beta \left( 2m^2_{H^0_3}-M^2_1 \right) \right],
\nonumber \\
&& \lambda_5={2\over v^2} \left( M^2_1-m^2_{H^0_3} \right),
 \end{eqnarray}
where $c_\theta$ and $s_\theta$ are abbreviations for $\cos\theta$ and $\sin\theta$ for $\theta = \alpha, \beta$, respectively, and $M_1$ and $M_2$ are defined as 
\begin{eqnarray}
M_1^2 = -\frac{v}{\sqrt{2} c_\beta}\mu_1 ~,~ \ \ M_2^2 =-3\sqrt{2} c_\beta v \mu_2 ~.
\end{eqnarray}

The Higgs boson trilinear self coupling in the model is therefore modified approximately as
\begin{eqnarray}
g_{hhh}\simeq \left\{ 1-{\mu_1^2v^2\over m_2^4} 
\left[ {7\over 8} -{3\over 2} {v^2\over m^2_h}
\left( (2\lambda_4+\lambda_5)+{\mu_1\mu_2\over m^2_2} 
\right) \right]
\right\} g_{hhh}^{\rm SM} ~,
\end{eqnarray}
where $g_{hhh}^{\rm SM}$ denotes the SM Higgs triple coupling shown in Eq.~(\ref{g_SM}). 
On the other hand, the coupling between one $H^0_1$ and two $h$ is
\begin{eqnarray}
g_{H^0_1hh}=&&
24\lambda_1 c^2_\alpha s_\alpha v_\phi
+ 2 \left[ \sqrt{3} c_\alpha v_\Delta (3c^2_\alpha -2)
+ s_\alpha v_\phi (1-3c_\alpha^2) \right]
(2\lambda_4+\lambda_5) 
\nonumber \\
&& 
+ 8\sqrt{3} c_\alpha s_\alpha^2 v_\Delta (\lambda_3+3\lambda_2)
+ {\sqrt{3}\over 2} \mu_1 c_\alpha (3c_\alpha^2 -2)
+ 4\sqrt{3} \mu_2 c_\alpha s^2_\alpha ~. \nonumber
\end{eqnarray}
Couplings of neutral Higgs bosons to fermions and gauge bosons relevant to this analysis are expressed in terms of the corresponding SM values as:
\begin{align}
\begin{split}
&
g_{hf\bar{f}} = {c_\alpha \over s_\beta} g_{hf\bar{f}}^{\rm SM} ~,
\qquad
g_{hVV} = \left(s_\beta c_\alpha - \sqrt{8\over 3} c_\beta s_\alpha \right)g_{hVV}^{\rm SM} ~,
\\
&
g_{H_1^0f\bar{f}} = {s_\alpha \over s_\beta} g_{hf\bar{f}}^{\rm SM} ~,
\qquad
g_{H_1^0VV} = \left( s_\beta s_\alpha + \sqrt{8\over 3} c_\beta c_\alpha \right)g_{hVV}^{\rm SM} ~.
\end{split}
\end{align}

\section{Higgs boson pair production }
\label{sec:hhprod}

\begin{figure}[t]
\centering
\includegraphics[width=3in]{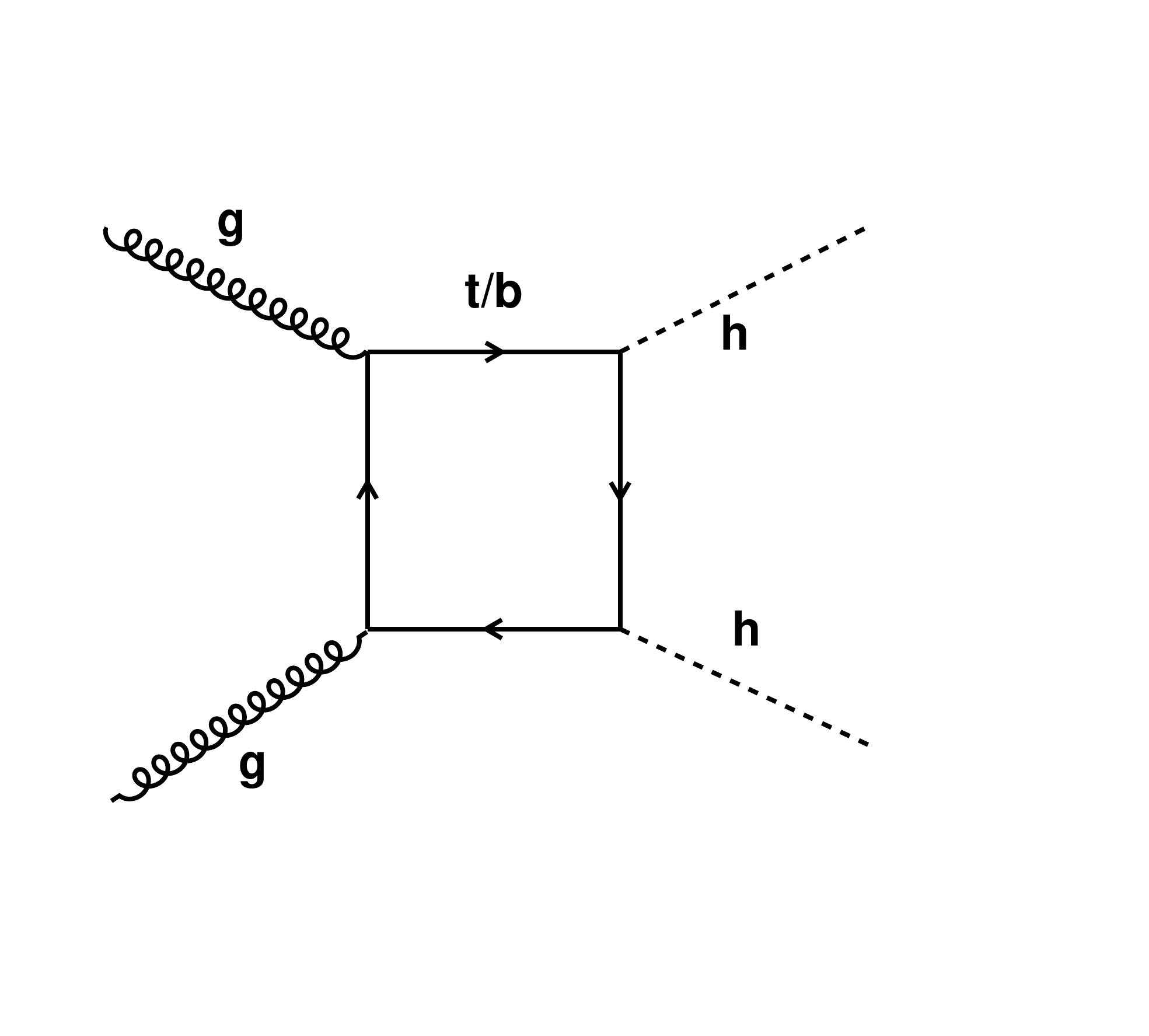} 
\includegraphics[width=3in]{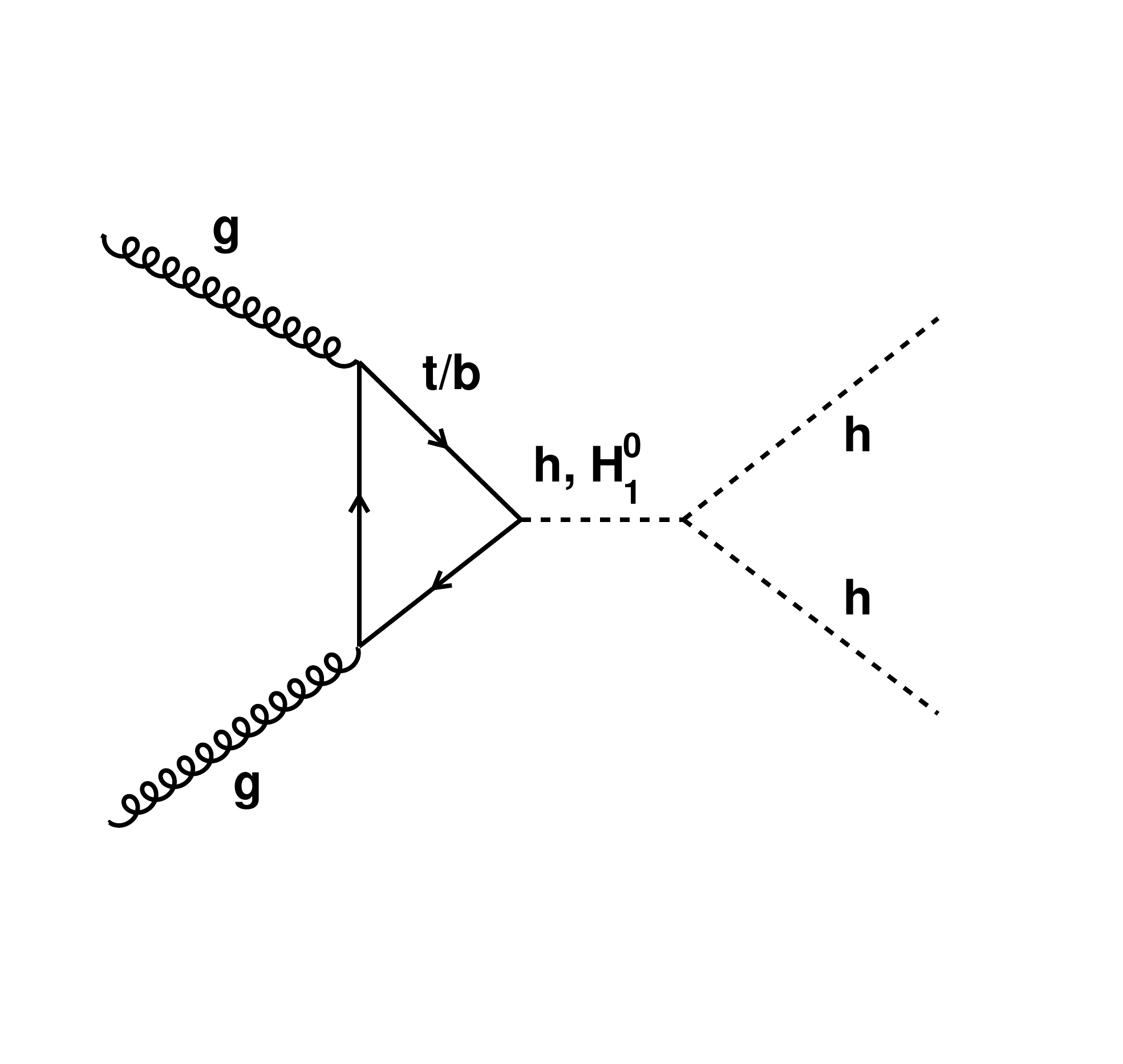}
\\
\vspace{-15mm}
(a) \hspace{75mm} (b)
\caption{\label{FR} Feynman diagrams for Higgs bosons pair production in the GM model. 
}
\end{figure}

As shown in Fig.~\ref{FR}, SM-like Higgs boson pair production in the GM model at the LHC receives contributions from both non-resonant process (plot (a)), mainly through top and bottom quark loops, and resonant process through the heavy $H_1^0$ decay (plot (b)).
The differential cross section for the process $g(p_1) g(p_2) \rightarrow h(p_3) h(p_4) $ is given by~\cite{plehn}
\begin{eqnarray}
\frac{d\hat\sigma(gg\to hh)}{d\hat{t}}
=&
\displaystyle
\frac{G_F^2 \alpha_s^2}{512(2\pi)^3}
\left[ \left| \lambda_{hhh}\kappa_{F_h} D(\hat{s}) F_\triangle
	+ \lambda_{H^0_1hh}\kappa_{F_{H^0_1}} \bar{D}(\hat{s}) F_\triangle
	+ \kappa^2_{F_h}F_\Box \right|^2 + \left| \kappa^2_{F_h}G_\Box \right|^2
\right] ~,
\nonumber
\\
& 
\mbox{with}~
\displaystyle
D(\hat{s})= \frac{3m_h^2}{\hat{s}-m_h^2+im_h\Gamma_h}
~,~
\bar{D}(\hat{s})= \frac{3m_h^2}{\hat{s}-m_{H^0_1}^2+im_{H^0_1}\Gamma_{H^0_1}} ~,
\end{eqnarray}
where $\kappa_{F_{h}}=g_{hf\bar{f}}/g^{SM}_{hf\bar{f}}$, $\kappa_{F_{H^0_1}}=g_{H^0_1f\bar{f}}/g^{SM}_{hf\bar{f}}$, $\lambda_{hhh} = g_{h hh}/g^{SM}_{hhh}$, $\lambda_{H^0_1hh} = g_{H^0_1 hh}/g^{SM}_{hhh}$ and
$\hat{s}=(p_1+p_2)^2$,
$\hat{t}=(p_1-p_3)^2$, and
$\hat{u}=(p_2-p_3)^2$ with $p_1+p_2=p_3+p_4$. 
The loop functions $F_\triangle$, 
$F_\Box$, and $G_\Box$ are given in Appendix A.1 of Ref.~\cite{plehn}. 
More explicitly,
\begin{eqnarray}
\frac{d\hat\sigma(gg\to hh)}{d\hat{t}}
\propto&&
\lambda_{hhh}^2|D(\hat{s})|^2[|F_\triangle|^2\kappa^2_{F_h}]+\lambda_{H^0_1hh}^2|\bar{D}(\hat{s})|^2[|F_\triangle|^2\kappa^2_{F_{H^0_1}}]\nonumber \\
&&+2\lambda_{hhh}\lambda_{H^0_1hh}\kappa_{F_h}\kappa_{F_{H^0_1}}\real(D(\hat{s})\bar{D}(\hat{s}))|F_\triangle|^2\nonumber \\
&&+2[\lambda_{hhh}\kappa_{F_h}^3 \real(D(\hat{s}) F_\triangle F^*_\Box)+\lambda_{H^0_1hh}\kappa_{F_{H^0_1}} \kappa_{F_h}^2 \real(\bar{D}(\hat{s}) F_\triangle F^{*}_\Box)]\nonumber\\
&&+ [ |F_\Box|^2+|G_\Box|^2 ] \kappa_{F_h}^4 ~.
\label{eq:explicit}
\end{eqnarray}
In the following, we will focus in the scenario where $m_{H_1^0} \gtr 2m_h$ and a pair of SM-like Higgs bosons can be produced via the production and decay of $H_1^0$.  In this case, we divide the total cross section into resonant and nonresonant contributions.  For the resonant production of the Higgs boson pair, we employ the narrow width approximation and calculate the production cross section of $H_1^0$, $\sigma(g g \rightarrow H_1^0)$, times its decay branching ratio to two Higgs bosons, $BR(H^0_1\rightarrow h h)$.  
Consider the dominant $H^{0}_{1}$ production by GGF at the LHC~\footnote{Here and the following, we tacitly consider only the dominant GGF production mechanism.  The vector boson fusion production mechanism is generally smaller by one order of magnitude~\cite{baglio, loop}.  This also makes our later production rate estimates more conservative.}.  Since the production of $H_1^0$ takes the same form as the SM Higgs boson production, the production cross section can be obtained by rescaling the result of SM Higgs boson with the modified Yukawa couplings and different masses.  We then have the resonant production of Higgs boson pairs as
\begin{align}\label{xsec_r}
\sigma(p p \rightarrow H^0_1 \rightarrow h h ) 
= \sigma(g g \rightarrow h)_{m_h \to m_{H_1^0}}\times \kappa_{F_{H_1^0}}^2\times BR(H^{0}_{1} \rightarrow h h) ~.
\end{align}

In view of the scaling of couplings in different parts of Eq.~(\ref{eq:explicit}), the nonresonant production cross section of a pair of Higgs boson can be parameterized as 
\begin{align}\label{xsec_nr}
\sigma(gg \to hh)
=&
\sigma_{\rm SM}(gg\to hh)
\Big[
\lambda_{hhh}^2\kappa^2_{F_h}c_1(s)
+ \lambda_{hhh}\kappa_{F_h}^3c_2(s)
+ \kappa_{F_h}^4c_3(s)
\nonumber \\
&
\qquad
+ \lambda_{hhh}\lambda_{H^0_1hh}\kappa_{F_h}\kappa_{F_{H^0_1}}c_4(s)
+ \lambda_{H^0_1hh}\kappa_{F_{H^0_1}} \kappa_{F_h}^2\bar{c}_2(s)
\Big] ~,
\end{align}
where we have removed the $H_1^0$ resonant production channel from the above expression to avoid double counting with Eq.~(\ref{xsec_r}). The coefficients $c_1 = 0.263$, $c_2 = -1.310$, $c_3 = 2.047$, and $c_4 = -0.001$ for $\sqrt{s} = 13$~TeV.  We also take a good approximation that $\bar{c}_2 = c_2$ when the production is off the resonance.  Our estimates of resonant production cross section to be given in the next section are scaled from the GGF single Higgs boson production cross section calculated at NNLO+NNLL QCD+NLO EW~\cite{deFlorian:2016spz}.  The SM Higgs boson pair production appearing in Eq.~(\ref{xsec_nr}) is calculated at NLO~\cite{Borowka:2016ehy}.

In this work, we use GMCALC~\cite{GMCALC} to calculate the Higgs mass spectrum, couplings and branching ratios in the GM model. Both theoretical and experimental constraints are taken into account, including tree-level unitarity, stability of Higgs potential, check of electroweak vacuum, and data of $ b\rightarrow s \gamma$ and $B^0_s\rightarrow \mu^+ \mu^-$ decays.  We have scanned 140,000 points in the parameter space of $ -90^\circ < \alpha <90^\circ$, $0<v_\Delta <60~{\rm GeV}$ and  $m_{H^0_1} \lesssim 1000~{\rm GeV}$.  We find that in a restricted region in the $\alpha$-$v_\Delta$ plane $m_{H^0_1}$ can be as heavy as 1~TeV, while most other space allows a maximum of around $700$~GeV.  It is a general feature that as $H^0_1$ becomes heavier, the range of $BR(H^0_1\rightarrow h h)$ becomes narrower and closer to 1, meaning that a heavy $H^0_1$ preferentially decays to a pair of SM-like Higgs bosons.

\begin{figure}[t]
\centering
\includegraphics[width=3in]{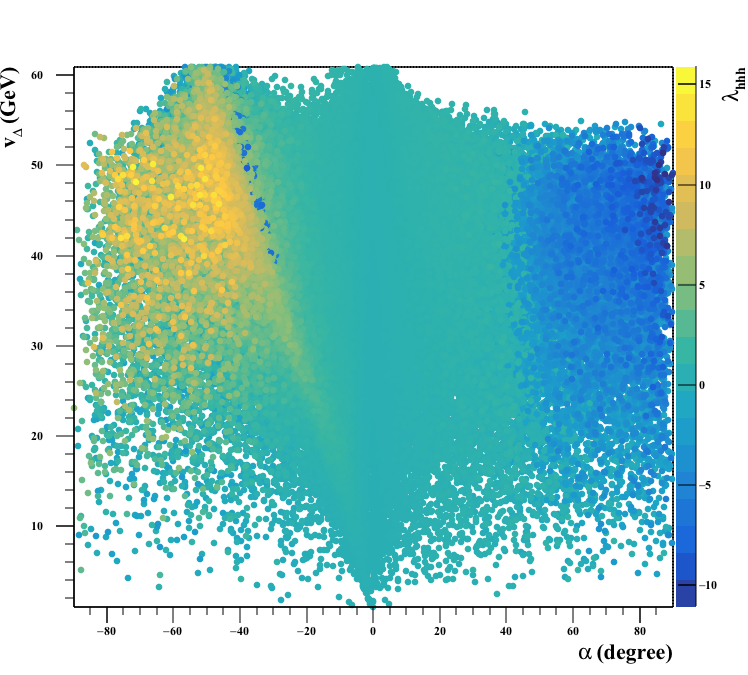} ~~
\includegraphics[width=3in]{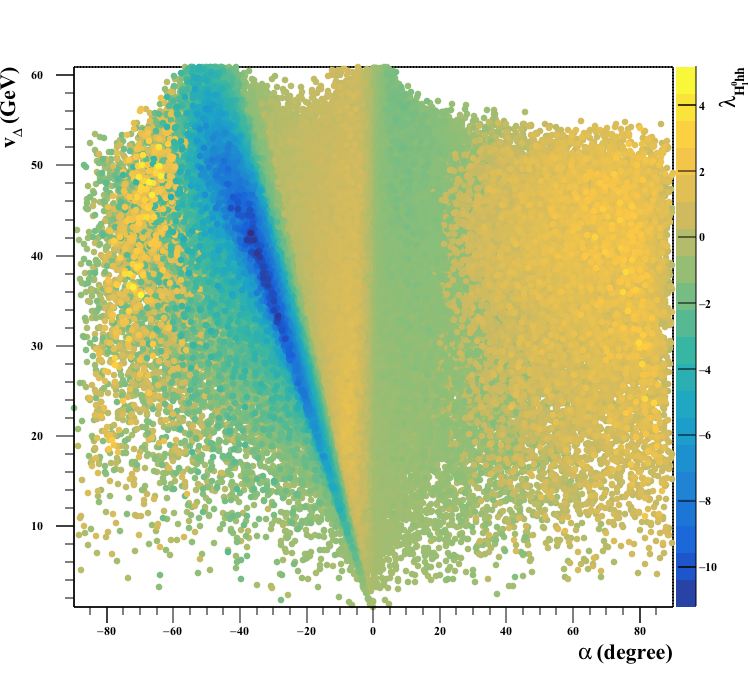}
\\
\vspace{-3mm}
(a) \hspace{7.5cm} (b)
\\
\includegraphics[width=3in]{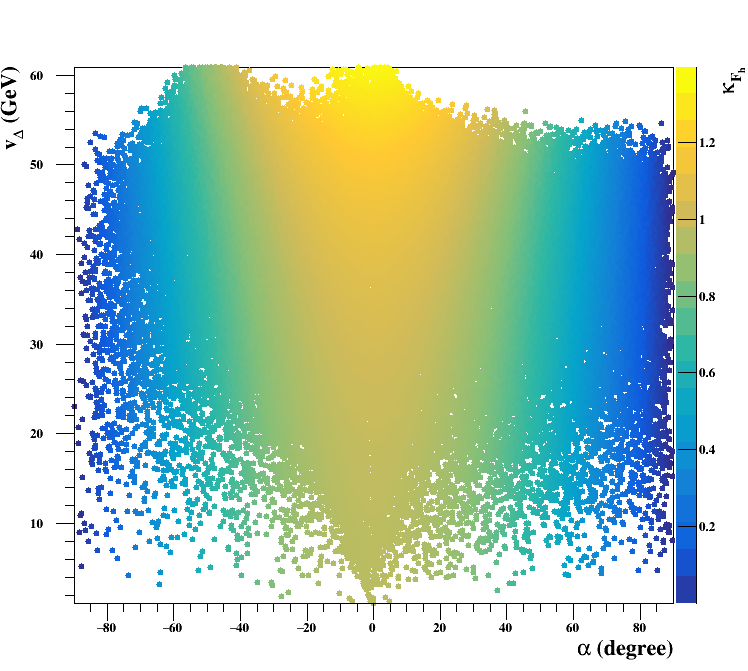} ~~
\includegraphics[width=3in]{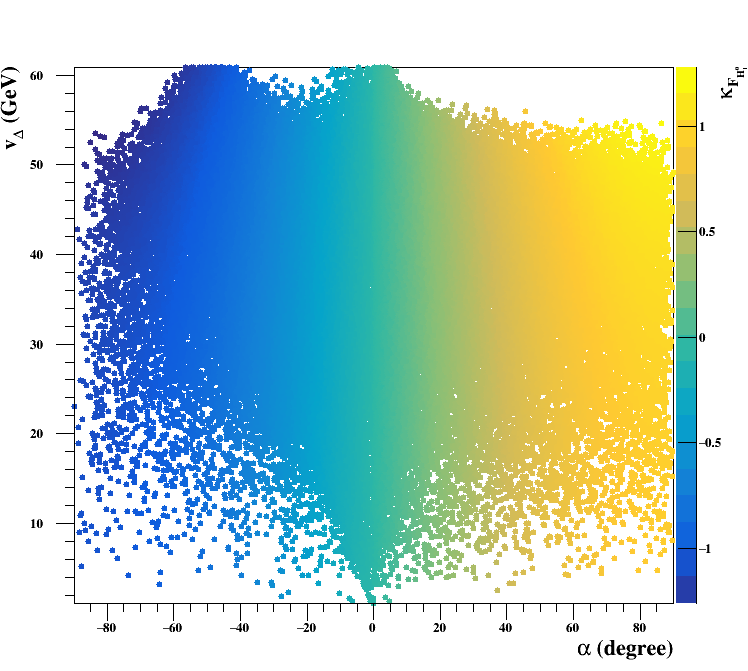}
\\
\vspace{-3mm}
(c) \hspace{7.5cm} (d)
\caption{\small \label{GM_ps1}  Couplings of $h$ and $H_1^0$ in the $\alpha$-$v_\Delta$ plane, with $m_{H_1^0} > 125$~GeV.  Plots (a) and (b) show respectively $\lambda_{hhh}$ and $\lambda_{H_1^0hh}$ with maximally allowed absolute value.  Plots (c) and (d) give respectively $\kappa_{F_{h}}$ and $\kappa_{F_{H_1^0}}$.
}\end{figure}

Fig.~\ref{GM_ps1} shows the couplings of $h$ and $H_1^0$.  Since each point in the $\alpha$-$v_\Delta$ plane allows certain ranges of $\lambda_{hhh}$ and $\lambda_{H_1^0hh}$, we show in plots (a) and (b) only those with the maximal absolute values.  As shown in the plots, $\lambda_{hhh}$ varies roughly in the range of $-20$ to $20$, $\lambda_{H^0_1hh}$ varies roughly between $-12$ and 6, $\kappa_{F_{h}}\lesssim 1.2$, and $|\kappa_{F_{H_1^0}}| \lesssim 1$.  In the plots of $\lambda_{hhh}$ and $\lambda_{H^0_1hh}$, one can clearly see a region (roughly from the origin to $\alpha \sim -40^\circ$ and $v_\Delta \sim 50$~GeV) in which both couplings attain large absolute values.  In particular, when $\lambda_{hhh}$ is negative (or $\lambda_{H^0_1hh}$ is positive), constructive interference between the box and triangle Feynman diagrams in Fig.~\ref{FR} would occur for that coupling and, in addition to the resonance effect, result in larger Higgs boson pair productions.

If $H^0_1$ is lighter than twice of SM-like Higgs boson mass, $m_{H_1^0} \lesssim 2m_h$, or the decay branching ratio of $H^0_1$ into two $h$'s is small, $BR(H^0_1\rightarrow h h)\sim 0$, the non-resonant production cross section, given by Eq.~(\ref{xsec_nr}), becomes more important and can be either enhanced or reduced in comparison with the SM prediction.

\begin{figure}[t]
\centering
\includegraphics[width=3in]{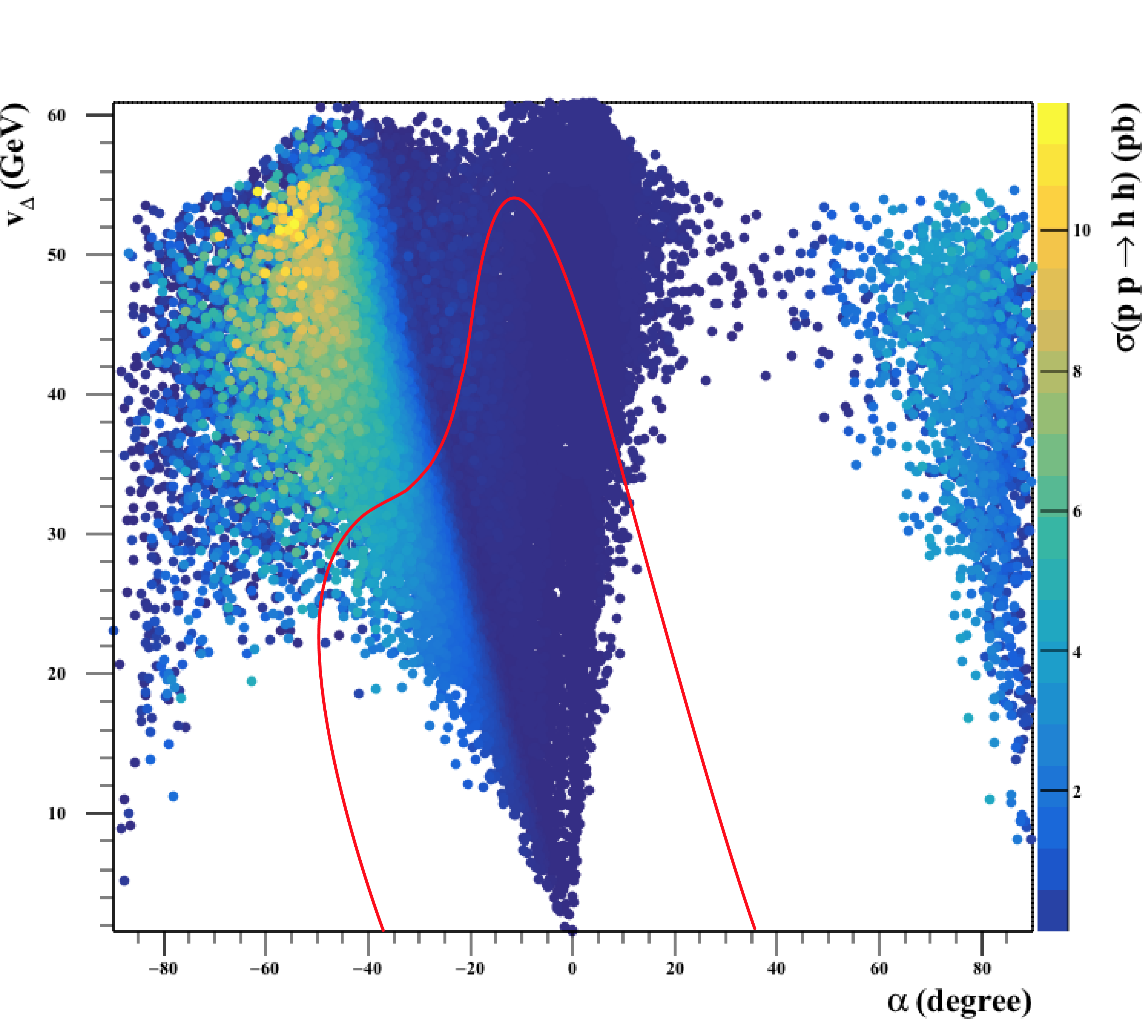} ~~
\includegraphics[width=3in]{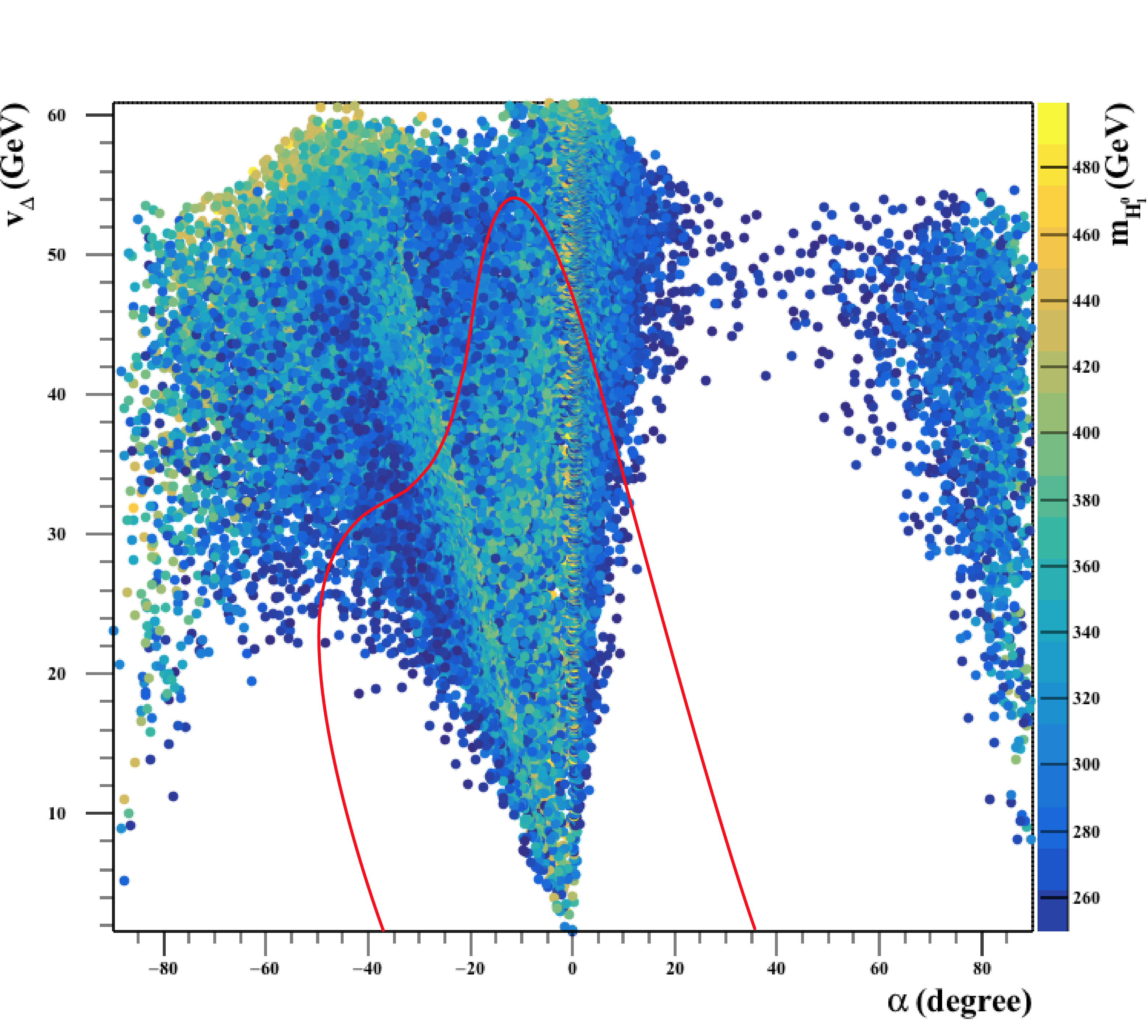}
\caption{\small \label{Hhh_BR} 
Maximum production cross section (left) and the corresponding $m_{H^0_1}$ (right) in the $\alpha$-$v_\Delta$ plane, assuming $BR(H^{0}_{1} \rightarrow h h) > 0$ and $m_{H^{0}_{1} } > 250$~GeV.}
\end{figure}

In Fig.~\ref{Hhh_BR}, we show the maximum resonant production cross section $\sigma(p p \rightarrow H^0_1 \rightarrow h h )$ (left plot) and the corresponding $m_{H^0_1}$ (right plot) in the $\alpha$-$v_\Delta$ plane.  Here we have further imposed the condition that $m_{H_1^0} > 2 m_h$ so that the $H_1^0 \to hh$ decay is kinematically allowed, resulting in fewer points in the parameter space than Fig.~\ref{GM_ps1}.
More scattered points accumulate in the region of $\alpha < 0$, and the maximum of cross section can reach about 6~pb within the red contour (for $\alpha \sim - 30^\circ$ and $v_\Delta \sim 30$~GeV).

\section{ Numerical results and direct searches constraints}
\label{sec:result} 

In this section, we select eight benchmark points on the $(\alpha,v_\Delta)$ parameter plane, chosen within the $2 \sigma$ bound from the Higgs data given in Ref.~\cite{Chiang:2015amq}: $(10,30)$, $(-10,50)$, $(-10,20)$, $(-30,20)$, $(-40,30)$, $(-45,20)$, $(-28,33)$ and the close-to-decoupling limit $(-1,1)$.  Here and afterwards, $\alpha$ and $v_\Delta$ are in units of degree and GeV, respectively.
The coupling scale factors and ranges of $m_{H^0_1}$ and $BR(H^0_1\rightarrow h h)$ for these benchmark points are listed in Table~\ref{benchmark point_set}.
Most benchmark points are located outside the heavy $m_{H_1^0}$ region, and $m_{H^0_1}\lesssim 500~{\rm GeV}$.  Only benchmark points C and G predict that $m_{H_1^0}$ can be as heavy as $\sim 1~{\rm TeV}$.
Note that the couplings of $H_1^0$ to quarks, $\kappa_{F_{H^0_1}}$, are larger in magnitude for benchmark points D, E, F and G.  Combined with the sizeable decay branching ratio of $H_1^0 \to h h$, the resonant production of SM-like Higgs boson pair can be significant.  In the close-to-decoupling limit, $(\alpha,v_\Delta)=(-1, 1)$, the pair production of $h$ becomes virtually the same as the SM prediction. 

{\squeezetable
\begin{table}[t]
\begin{tabular}{c| c c c c c c c c}
\hline\hline
benchmark point & A & B & C & D & E & F & G & H \\
\hline
$(\alpha,v_\Delta)$ 
& $(10,30)$ & $(-10,50)$ & $(-10,20)$ & $(-30,20)$ 
& $(-40,30)$ & $(-45,20)$ & $(-28,33)$ & $(-1,1)$ \\
\hline
$\kappa_{F_h}$ & 1.049 & 1.204 & 1.012 & 0.889 & 0.816 & 0.727 & 0.954 & 0.999 \\
$\kappa_{F_{H^0_1}}$ & 0.185 & $-0.212$ & $-0.178$ & $-0.514$ & $-0.685$ & $-0.727$ & $-0.507$ & $-0.018$ \\
$\kappa_{V_h}$ & 0.827 & 0.969 & 1.024 & 1.031 & 1.081 & 0.954 & 1.108 & 1.00 \\
$\kappa_{V_{H^0_1}}$ & 0.718 & 0.782 & 0.201 & $-0.161$ & $-0.172$ & $-0.423$ & 0.113 & $1.32\times 10^{-3}$ \\
$m_{H^0_1}$ & 250--301 & 250--455 & 250--954 & 250--315 & 250--402 & 250--273 & 250--1373 & 250--492 \\
$BR(H^0_1\rightarrow h h)$ & 0.004--0.16 & 0.0014--0.133 & 0.009--0.186 & 0.244--0.954 & $2\times10^{-4}$--0.96 & $2\times10^{-5}$--0.5 & $7\times10^{-3}$--0.81 & 0.6--0.99 \\
\hline\hline
\end{tabular}
\caption{\label{benchmark point_set}  
Coupling scale factors, the range of $m_{H^0_1} ~(\gtrsim 2m_h)$ and the range of $BR(H^0_1\rightarrow h h)$ for 8 benchmark points.  We have scanned 3000 points for each benchmark point set, where $\alpha$ is in units of degree and $v_\Delta$ and $m_{H^0_1}$ are in units of GeV.}
\end{table}}

\begin{figure}[t]
\centering
\includegraphics[width=3in]{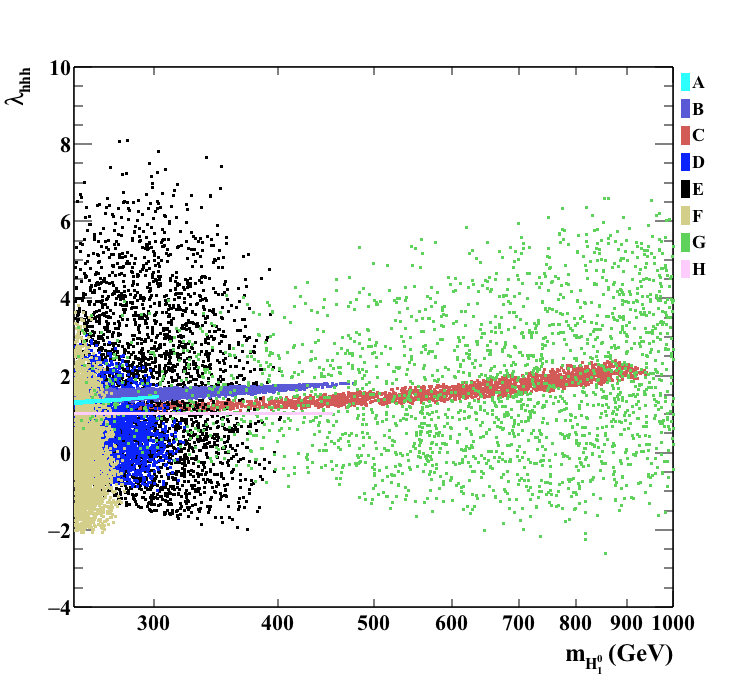}
\includegraphics[width=3in]{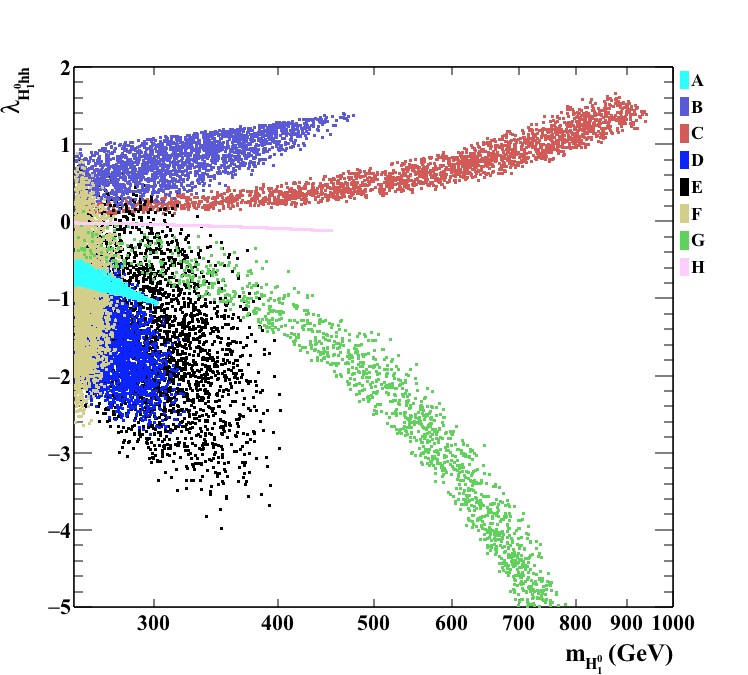}
\caption{\small \label{hhh_mH}Scatter plots of scalar couplings $\lambda_{h hh}$ (left) and $\lambda_{H^{0}_{1} hh}$ (right) as a function of $m_{H_1^0}$. 
}\end{figure}

In addition to the couplings that are fixed by the chosen values of $(\alpha,v_\Delta)$ shown in Table~\ref{benchmark point_set}, the scalar self-couplings are also crucial for the production of $hh$ pairs.  We show in Fig.~\ref{hhh_mH} the scatter plots of $\lambda_{h hh}$ (left plot) and $\lambda_{H^{0}_{1} hh}$ (right plot) for each benchmark point.  The trilinear self-coupling of $h$ can significantly deviate from the SM value, and even flip its sign in benchmark points  D, E, F and G, resulting in a wide range of possible values. 
For the coupling of $H_1^0$ to two light Higgs bosons $h$, benchmark points A, D, E, F, and G predicts values with an opposite sign to the SM Higgs self-coupling, with the latter four having particularly wide ranges.  Only benchmark points B and C predict a positive sign and $\sim {\cal O}(1)$ for the coupling.

Before presenting our simulations, let us summarize the current situation of the search for Higgs boson pairs at the LHC.  Here we only focus on the $bb\gamma\gamma$ and $4b$ final states since these two channels impose stronger constraints and are complementary when a resonance $H_1^0$ exists.  The $bb\gamma\gamma$ channel serves as a good search channel in the lower mass regime as it has a cleaner signature, particularly for the non-resonant Higgs boson pair production in the SM.  In the case of resonant production via a heavy resonance ($M_X \gtrsim 500$~GeV), its efficiency becomes lower than the $4b$ channel.  This is because the photon pair coming from the more boosted Higgs boson decay will be very collinear.  Experimentally, separating the two photons in this case significantly lowers the efficiency.

At ATLAS, the search for a light $H_1^0$ with mass $275~{\rm GeV} \leq m_{H^0_1}\leq 400~{\rm GeV}$ is constrained by the $bb\gamma\gamma$ channel~\cite{atlas_13,bbaa_hh}.  The efficiencies for signal events to pass the selection criteria are about $5-8\%$, depending on the mass of $H_1^0$.  It is shown that the distribution of invariant mass of the $h$ pair, $M_{hh}$, in the SM peaks around $400$ GeV at the LHC~\cite{hh-sm}, and the peak position does not shift much as the collision energy varies from 8~TeV to 100~TeV.  Therefore, a light resonant can contribute to the $h$ pair production rate through both interference effect and on-shell production.

The $4b$ search channel used by the ATLAS Collaboration~\cite{atlas_13_2,4b_hh}, on the other hand, gives a cross section upper limit for a heavy scalar resonance in the mass range of $500 \ {\rm GeV} \leq m_{H^0_1}\leq 1000\ {\rm GeV}$ using the resolved analysis, and $1000 \ {\rm GeV} \leq m_{H^0_1}\leq 3000\ {\rm GeV}$ using the boosted analysis.  The event selection efficiencies in the resolved analysis, where different cuts are applied for different masses of heavy resonance, are given by
{\[
\begin{tabular}{c | c c c c c c}
Mass (GeV) & 500 & 600 & 700 & 800 & 900 & 1000\\
\hline
Efficiency~\cite{atlas_13_2} &0.95$\%$&1.91$\%$&2.55$\%$&2.86$\%$&3.14$\%$&3.45$\%$\\
\end{tabular}\]}
Here the calculation of efficiency assumes a 100\% branching ratio for the heavy scalar resonance to a pair of SM-like Higgs bosons and a fixed total decay width of 1~GeV.

In our simulations, events of Higgs boson pair production are generated with the loop-induced mode in {\tt Madgraph5 aMC@NLO}~\cite{Alwall:2014hca} with $m_h=125$~GeV.  The model file is adopted from the model database of {\tt FeynRules}~\cite{GM_MG5:NLO, FR}.
The decays of Higgs boson into $b\bar{b}$ and $\gamma\gamma$ are performed with {\tt MadSpin}~\cite{spin}.  The events are then passed to {\tt Pythia8}~\cite{Sjostrand:2007gs} for parton showering and hadronization, and the fast detector simulation in {\tt Delphes3} (ATLAS settings)~\cite{delphes3} is used to include the detector effects.  Finally, events are analyzed with {\tt MadAnalysis5}~\cite{MA5}.

\begin{figure}[t]
\includegraphics[width=3in]{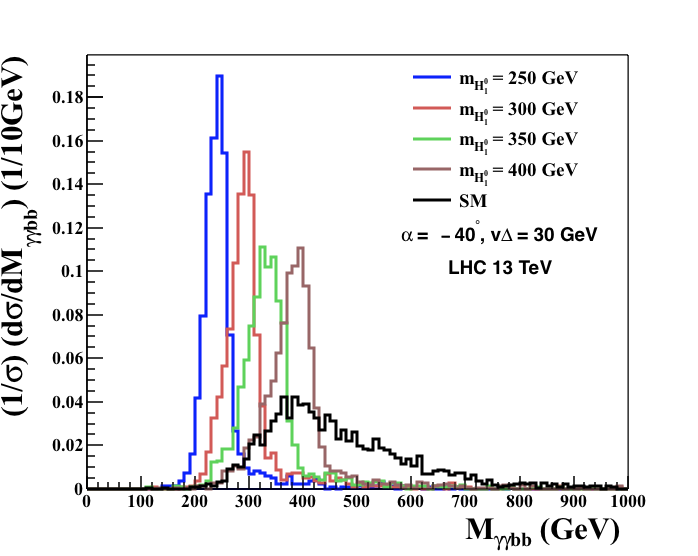} 
\includegraphics[width=3in]{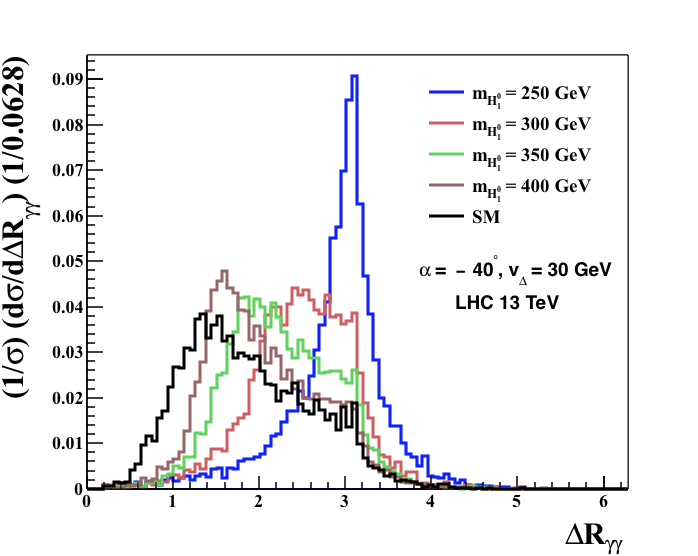}
\\
(a) \hspace{75mm} (b)
\\
\includegraphics[width=3in]{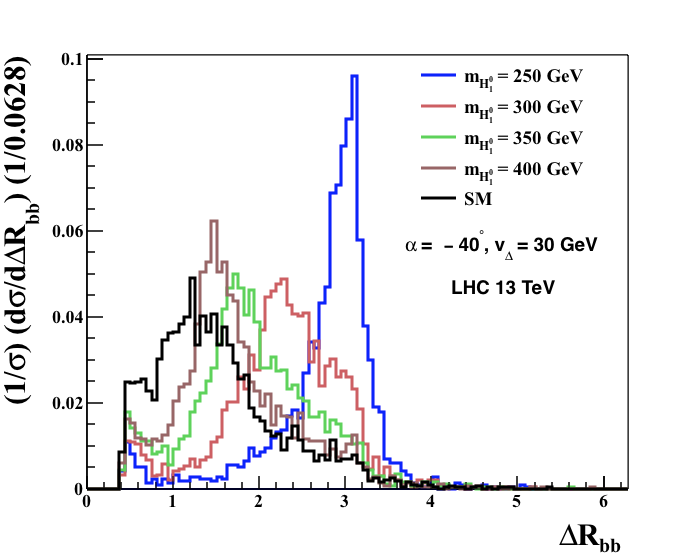}
\\
(c)
\caption{\small \label{kin_m4030} Kinematic distributions of the $bb\gamma\gamma$ channel for (a) the invariant mass $M_{\gamma\gamma b b}$, (b) the opening angle $\Delta R_{\gamma\gamma}$ and (c) the opening angle $\Delta R_{b b}$ for benchmark point E with different $m_{H_1^0}$ in comparison with the SM expectations at the 13-TeV LHC.
}
\end{figure}

In the case of light $H_1^0$ in the mass range $250~{\rm GeV} \le m_{H^0_1} \le 500~{\rm GeV}$, we follow the cuts used in the ATLAS $bb\gamma\gamma$ channel analysis~\cite{atlas_13}:
\begin{eqnarray}\label{ac_a}
&&N_\gamma \ge 2,\ N_b = 2 ~,\ P_T(j)> 25~{\rm GeV} ~,\ P_T(b)^{\rm lead, subl}>55,~ 35~{\rm GeV} ~,\nonumber \\
&& 105~{\rm GeV} < M_{\gamma \gamma} < 160~{\rm GeV},\ 95~{\rm GeV} < M_{b b} < 135~{\rm GeV} ~.
\end{eqnarray}
Here and the following, $N_p$ refers to the number of particle $p$, $P_T(h)$ is the transverse momentum of particle or system $h$, the superscripts ``lead'' and ``subl'' denote respectively the leading and subleading jets, and $M_{xx}$ ($x = b,\gamma$) is the invariant mass of the system.
The kinematic distributions in the invariant mass $M_{\gamma\gamma b b}$ and the opening angles $\Delta R$ of the two photons and of two $b$ jets are shown in Fig.~\ref{kin_m4030}, where we illustrate with different masses of $H_1^0$ in benchmark point E.  Unlike the broad invariant mass distributions peaked around 400~GeV in the SM, a clear resonance at the mass of $H_1^0$ can be readily identified in plot~(a).  The opening angle of the Higgs decay products $\Delta R \approx 2m_h/P_{T}(h)$, where $P_{T}(h)$ denotes the transverse momentum of the decaying $h$.  Since the production of Higgs boson pair via a lighter resonance generally has less boosted $h$, the opening angle of the Higgs decay products tends to be wider in this case, as seen in both plots~(b) and (c) of Fig.~\ref{kin_m4030}.  It is also noted that the reason for the SM background to have smaller $\Delta R$ in these two plots is because the Higgs pair production mainly comes from the non-resonance production ({\it i.e.}, the box diagram) that produces more Higgs bosons with larger $p_T$.

\begin{figure}[t]
\centering
\includegraphics[width=3in]{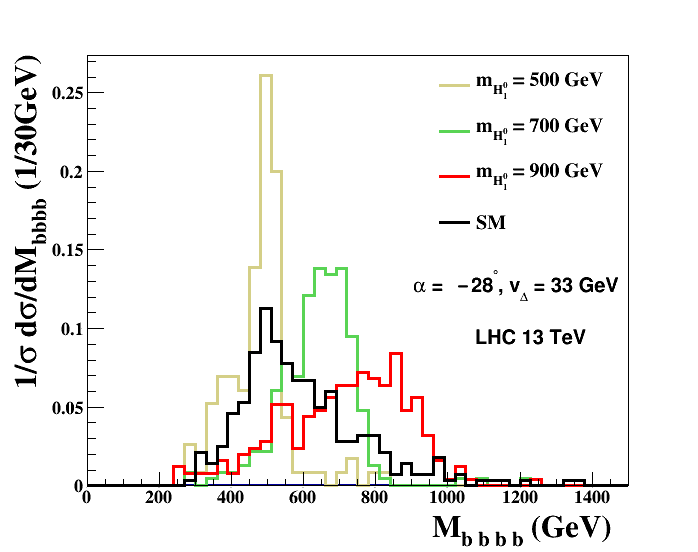} ~~
\includegraphics[width=3in]{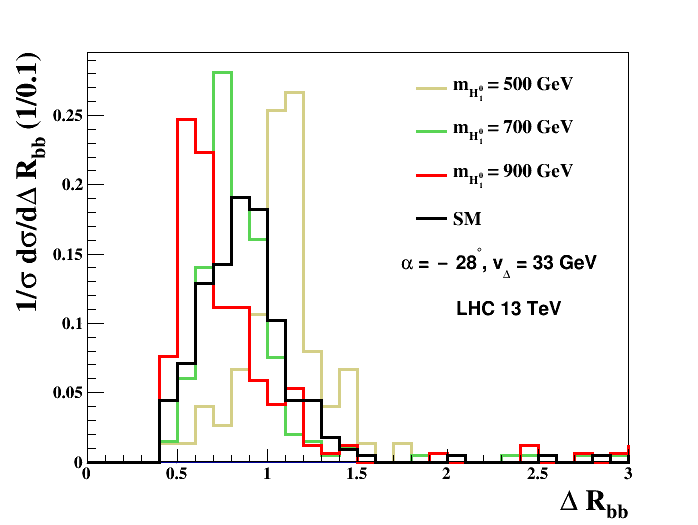}
\end{figure}

In the case of heavy $H_1^0$ with mass larger than $500$~GeV, the ATLAS $4b$ search using the resolved analysis is employed.  We take benchmark point G as an example to show the distribution in the invariant mass $M_{bbbb}$ and that in $\Delta R$ of the second and third energetic $b$ jets. 
The curves in the plots are the results after imposing the preselection cuts used by ATLAS for the $4b$ channel analysis:
\begin{eqnarray}\label{ac_b}
&& N_b \ge 4 ~,\ |\eta(j)| < 2.5 ~,\ P_T(b)> 40~{\rm GeV} ~,\nonumber \\
&& \Delta R(jj) < 1.5,\
P_T(j j)^{\rm lead, subl}>200, 150~{\rm GeV} ~.
\end{eqnarray}
We observe that as $m_{H_1^0}$ becomes heavier, the peak in the distribution of $M_{bbbb}$ becomes broader as its total width gets bigger.  The $\Delta R$ distribution also moves to smaller values, as expected.
In order to make a comparison with experimental constraints measured by the ATLAS Collaboration, we further follow their analysis to impose the additional mass-dependent cuts in our numerical simulations:
\begin{align}\label{ac}
P_T^{\rm lead}(jj) > &
\begin{cases}
400~{\rm GeV} \qquad\qquad\qquad~ \mbox{if } M_{4j} > 910~{\rm GeV} ~,\\
200~{\rm GeV} \qquad\qquad\qquad~ \mbox{if } M_{4j} < 600~{\rm GeV} ~,\\
0.65M_{4j}-190~{\rm GeV} \qquad \mbox{otherwise} ~;\\
\end{cases}
\nonumber \\ 
P_T^{\rm subl}(jj) > &
\begin{cases} 
260~{\rm GeV} \qquad\qquad\qquad \mbox{if } M_{4j} > 990~{\rm GeV} ~,\\
150~{\rm GeV} \qquad\qquad\qquad \mbox{if } M_{4j} < 520~{\rm GeV} ~,\\
0.23M_{4j}+30~{\rm GeV} \qquad \mbox{otherwise} ~;\\
\end{cases}
\nonumber \\ 
|\Delta\eta(jj)| < &
\begin{cases} 
1.0 \qquad\qquad\qquad\qquad\quad \mbox{ if } M_{4j} < 820~{\rm GeV} ~,\\
1.6\times 10^{-3} M_{4j}-0.28 \qquad \mbox{otherwise} ~.
\end{cases}
\end{align}
\begin{table}[t]
\begin{tabular}{c | c c c c | c c c c c c | c}
\hline\hline
Benchmark point & \multicolumn{4}{c|}{E} & \multicolumn{6}{c|}{G} & SM
\\
\hline
$(\alpha, v_\Delta)$& \multicolumn{4}{c|}{$(-40^\circ,\ 30~{\rm GeV})$} 
& \multicolumn{6}{c|}{$(-28^\circ,\ 33~{\rm GeV})$} & \\
$m_{H^0_1}~{\rm (GeV)}$& 250 & 300 & 350 & 400 & 500 & 600 & 700 & 800 & 900 & 1000 &
\\
$\Gamma_{H_1^0}$ (GeV)&0.68&5.37&10.62&8.05&6.75&9.04&18.91&27.83&34.67&51.00
\\
$BR(H^0_1\rightarrow h h)$&0.82&0.954&0.955&0.76&0.57&0.45&0.62&0.66&0.65&0.71 &
\\
$\sigma(pp \to hh)_{13-{\rm TeV}}$ (pb)&3.62&3.28&3.32&2.68&0.56&0.25&0.18&0.11&0.11&0.078
\\
Efficiency &5.6$\%$&6.4$\%$&7.2$\%$&8.8$\%$&2.57$\%$&4.15$\%$&3.65$\%$&2.45$\%$&0.86$\%$&0.97$\%$ &9.2$\%$
\\
\hline\hline
\end{tabular}
\caption{\label{benchmark point_e} Mass of $H_1^0$, its total decay width, its decay branching ratio and production rate to a pair of SM-like Higgs bosons, and the selection efficiency for benchmark point E in the $\gamma\gamma b b$ channel, benchmark point G in the $4b$ channel, and SM in the $b b \gamma\gamma$ channel at the 13-TeV LHC.}
\end{table}

The efficiencies for different masses of $H_1^0$ and the decay branching ratio to $hh$ for benchmark points E and G are listed in Table~\ref{benchmark point_e}.  Here we choose the other parameters to maximize the resonant Higgs pair production rate via GGF (and thus the branching ratio of $H_1^0 \to hh$), whose value is also given in the table.  The efficiency for the $bb\gamma\gamma$ channel in the SM is also given for a comparison. 
The efficiency for our cases depends on both the mass of $H_1^0$, its production rate, and its branching ratio to a pair of SM-like Higgs bosons.  For the $bb\gamma\gamma$ channel in the lower mass regime, the experimental cuts are designed to be optimal for the non-resonant production that is peaked around 400~GeV.  Therefore, we find that the efficiency in benchmark point E reduces as $m_{H_1^0}$ becomes smaller.  For the $4b$ channel in the higher mass regime, on the other hand, the cuts are designed for resonant production and will cut away non-resonant events if $m_{H_1^0}$ is sufficiently large.

\begin{figure}[t]
\centering
\includegraphics[width=3in]{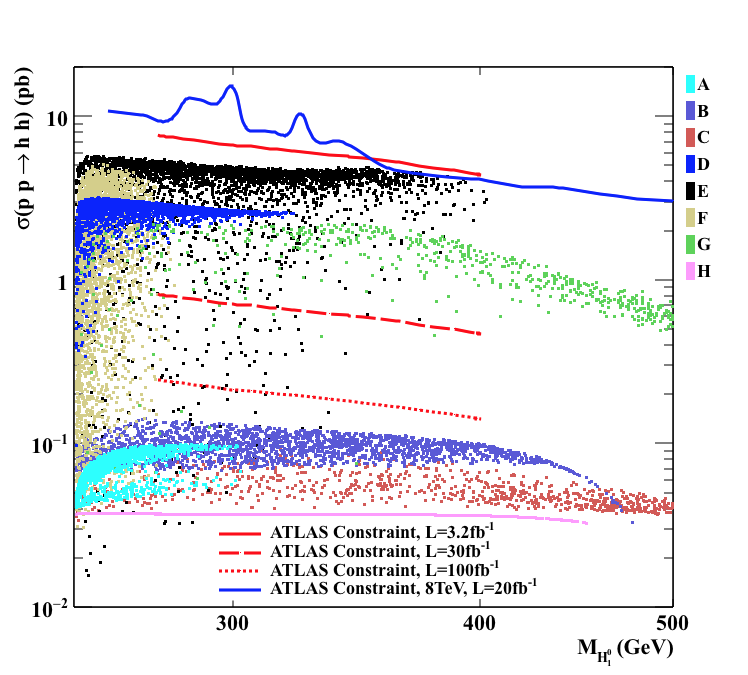} ~~
\includegraphics[width=3in]{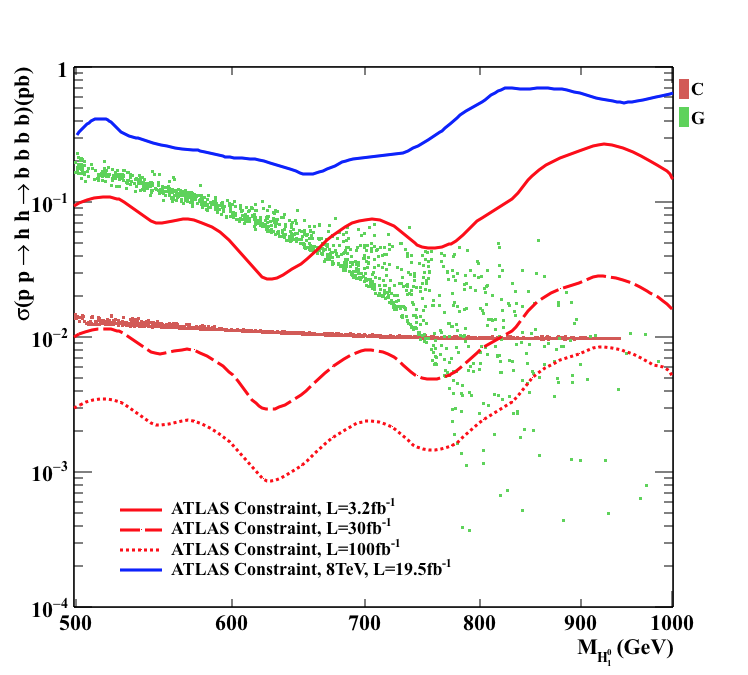}
\caption{\label{xsec}Estimated resonant cross section $\sigma(p p \rightarrow H^{0}_{1}\rightarrow h h)=\sigma(p p \rightarrow H_1^0) \times \kappa_{F_{H_1^0}}^2 \times BR(H^0_1 \rightarrow h h)$ versus $m_{H_1^0}$ for each benchmark point set at the 13-TeV LHC, with the luminosities of $3.2$~fb$^{-1}$ (red solid curves), $30$~fb$^{-1}$ (red dashed curves) and $100$~fb$^{-1}$ (red dotted curves).  The left plot is for the $\gamma\gamma b b$ channel in the lower mass regime, an the right plot is for the $4b$ channel in the higher mass regime.  Also shown are scaled constraints of the 8-TeV data (blue solid curves) with the luminosities of $20$~fb$^{-1}$ (left plot) and $19.5$~fb$^{-1}$ (right plot)}.
\end{figure}

Fig.~\ref{xsec} plots our estimates of Higgs pair production cross sections for the eight benchmark points, including both resonant and non-resonant contributions [from Eq.~(\ref{xsec_r}) and Eq.~(\ref{xsec_nr})].  For each benchmark point set, we have scanned 3000 points \footnote{Note that if we sample more points, the cross section ranges may only go slightly wider.}.  Most of the parameter space in benchmark points D, E, F, and G predict larger cross sections at the level of a few picobarns, in comparison with the other benchmark points.  This is because the Higgs boson trilinear coupling $g_{hhh}$ in these four benchmark points can go negative, resulting in a constructive interference between the box and triangle Feynman diagrams in Fig.~\ref{FR}.  It is noted that at the same time in these benchmark points, $g_{H_1^0 hh}$ is also negative, resulting in destructive interference to cancel part of the aforementioned constructive interference.  The left plot shows scattered points for all the benchmark points in the mass range of $250$~GeV $\le m_{H_1^0} \le 500$~GeV.  The right plot shows scattered points for benchmark points C and G in the mass range of $500$~GeV $\le m_{H_1^0} \le 1$~TeV as only they allow larger $m_{H_1^0}$ among the benchmark points considered here.

We also show the current constraints (red solid curves) on the searches for $H_1^0$ from the $\gamma\gamma b b$ channel~\cite{atlas_13} and the $4b$ channel~\cite{atlas_13_2} done by the ATLAS Collaboration using the $3.2~{\rm fb^{-1}}$ dataset at the 13-TeV LHC.  As a comparison, we also show the constraints (blue curves) of the corresponding searches from LHC Run-I \cite{atlas_8} after taking into account the acceptances and rescaling of the parton luminosity.  It is seen that benchmark point E is close to the constraint of the $\gamma\gamma b b$ channel.  The parameter space of $500~{\rm GeV}\lesssim M_{H_1^0}\lesssim 650~{\rm GeV}$ for benchmark point G is already excluded by the $4b$ channel search.  We also estimate the projected exclusion limits (red dashed curves for an integrated luminosity of $30$~fb$^{-1}$ and red dotted curves for $100$~fb$^{-1}$) when more data are collected.  With $30$~fb$^{-1}$, the LHC has the sensitivity to most of the parameter space with the $H_1^0$ mass heavier than twice the Higgs boson mass for benchmark points D, E, F and G.  The parameter space of heavier $H_1^0$ with mass larger than $500~{\rm GeV}$ for benchmark point C can be probed as well.

We note that the ATLAS $\gamma\gamma b b$ and $4b$ constraints are rescaled with the efficiencies for benchmark points E and G, respectively (see Table~\ref{benchmark point_e}).  Different benchmark points would have slightly different efficiencies.  In addition to the current luminosity of $3.2$~fb$^{-1}$ (drawn in red solid curves), we also plot those for $30$~fb$^{-1}$ (red dashed curves) and $100$~fb$^{-1}$ (red dotted curves).  Among the eight scenarios considered here, benchmark points E and G predict largest cross sections in the lower and higher mass regimes, respectively, and benchmark points C and G allow wider mass ranges for $H^{0}_{1}$.  The pink scattered points for benchmark point H have production rates approaching the SM prediction.

\section{Conclusion}
\label{sec:con}

In this paper, we have studied in the Georgi-Machacek (GM) model the SM-like Higgs boson pair production through the gluon-gluon fusion (GGF) process at the 13-TeV LHC.  We find that under various theory and experimental constraints, the Higgs boson couplings (self and with other SM particles) can have some deviations from the SM values.  In particular, the model and current data even allow an interesting possibility that the Higgs boson self-coupling $g_{hhh}$ can flip its sign from the SM value.  In addition, the existence of the heavier Higgs singlet $H_1^0$ in the model gives an additional contribution to the di-Higgs production cross section through its mixing with the SM-like Higgs boson.  The mass of $H_1^0$ can in some cases be as heavy as $1$~TeV, especially in some parameter region with a negative mixing angle $\alpha$.

When $H_1^0$ is sufficiently heavy to decay into a pair of SM-like Higgs bosons, the production rate can be significantly enhanced, particularly when the Higgs trilinear coupling $g_{hhh}$ becomes negative as constructive interference would occur.  We also note that at the same time the other Higgs trilinear coupling $g_{H_1^0hh}$ is also negative to result in a smaller destructive interference.  For illustration purposes, we select eight benchmark points and perform a detailed numerical study.  The Higgs boson pair production rate is estimated and compared with current and projected search bounds given by the ATLAS Collaboration.  A couple of scenarios considered here can be probed or ruled out by the LHC experiments in the near future.

\section*{Acknowledgments}  
The authors are grateful to J.~Baglio for pointing out useful references.
This work was supported in part by the Ministry of Science and Technology of Taiwan under Grant Nos.~MOST-105-2112-M-003-010-MY3 (CRC) and MOST-104-2628-M-002-014-MY4 (CWC).


\end{document}